\documentclass[sigconf,nonacm,balance=false]{acmart}
\setcopyright{none}
\renewcommand\footnotetextcopyrightpermission[1]{} % suppress copyright footnote

\usepackage{amsmath,amsthm}

\usepackage{amssymb}
\usepackage{booktabs}
\usepackage{graphicx}
\usepackage{microtype}
\usepackage{enumitem}
\usepackage{xspace}
\usepackage{float}
\usepackage{placeins}
\usepackage{cleveref}

\newtheorem{definition}{Definition}
\newtheorem{theorem}{Theorem}
\newtheorem{lemma}{Lemma}
\newtheorem{proposition}{Proposition}
\newtheorem{corollary}{Corollary}

\newtheorem{remark}{Remark}
\crefname{definition}{Definition}{Definitions}
\crefname{theorem}{Theorem}{Theorems}
\crefname{lemma}{Lemma}{Lemmas}
\crefname{proposition}{Proposition}{Propositions}
\crefname{corollary}{Corollary}{Corollaries}
\crefname{assumption}{Assumption}{Assumptions}
\crefname{remark}{Remark}{Remarks}
\crefname{equation}{Equation}{Equations}
\crefname{figure}{Figure}{Figures}
\crefname{table}{Table}{Tables}
\crefname{section}{Section}{Sections}

\newcommand{\creators}{\mathcal{C}}       % creator set
\newcommand{\viewers}{\mathcal{V}}         % viewer set
\newcommand{\nactive}{C}                    % number of active creators
\newcommand{\poolsize}{n}                   % addressable pool size (full market)
\newcommand{\util}{u}                       % match utility
\newcommand{\ordstat}[1]{\util_{(#1)}}      % order statistic
\newcommand{\loss}{L}                       % engagement loss
\newcommand{\pmin}{p^{\ast}}                % minimum viable traffic fraction
\newcommand{\suppmin}{m^{\ast}}            % minimum viable in-arm supply
\newcommand{\E}{\mathbb{E}}
\newcommand{\Prob}{\mathbb{P}}
\newcommand{\eqdef}{\triangleq}

\begin{document}

\title{The Price of Isolation: Estimating the Ecosystem Cost of Symmetric Two-Sided A/B Testing}

\author{Yuanyuan Shen}
\affiliation{\institution{Snap Inc.}\city{New York}\state{NY}\country{USA}}
\email{yshen2@snapchat.com}

\author{Yiren Yan}
\affiliation{\institution{Snap Inc.}\city{Palo Alto}\state{CA}\country{USA}}
\email{yyan5@snapchat.com}

\author{Wenjie Li}
\authornote{Work done while at Snap Inc.}
\affiliation{\institution{Snap Inc.}\city{Palo Alto}\state{CA}\country{USA}}
\email{wenjie.li@snapchat.com}

\author{Chunhui Zhu}
\affiliation{\institution{Snap Inc.}\city{Palo Alto}\state{CA}\country{USA}}
\email{chunhui.zhu@snapchat.com}

\begin{abstract}
On two-sided content platforms, symmetric two-sided isolation (assigning matched
fractions of creators and viewers to isolated treatment and control submarkets)
is widely used for creator-side
and cold-start experiments because it removes cross-arm marketplace interference.
Isolation, however, thins each viewer's candidate catalog, and intuition suggests
the resulting engagement cost should fade as the platform grows: a small fraction
of a vast catalog is still vast. We show that, in an order-statistics model of
engagement, whether this intuition holds depends on the upper tail of match
quality. Extreme-value theory yields tail-class loss laws with a sharp dichotomy:
for light or bounded tails the loss vanishes as the candidate pool grows, whereas
under heavy tails it converges to a size-independent constant, so expanding the
candidate pool, even by orders of magnitude, does not asymptotically eliminate
the cost. Evidence from two production experiments on a large-scale content
platform is consistent with this picture: a pure A/A traffic sweep reveals
a measurable, depth-graded engagement cost; a one-sided catalog ablation
independently shows that per-viewer thinning contributes to the loss; and a tail
index calibrated on the small exploration pool predicts an effect consistent with
the one observed in the far larger full-catalog ablation. Isolation thus carries
a price that experimenters should budget for, like any other cost. We give
practitioners a preflight procedure that estimates it before launch, sizes
traffic accordingly, and recommends a fallback design when the predicted cost
exceeds a chosen tolerance.

\end{abstract}

\begin{CCSXML}
<ccs2012>
  <concept>
    <concept_id>10002951.10003317.10003347.10003350</concept_id>
    <concept_desc>Information systems~Recommender systems</concept_desc>
    <concept_significance>500</concept_significance>
  </concept>
  <concept>
    <concept_id>10002950.10003648</concept_id>
    <concept_desc>Mathematics of computing~Probability and statistics</concept_desc>
    <concept_significance>300</concept_significance>
  </concept>
</ccs2012>
\end{CCSXML}
\ccsdesc[500]{Information systems~Recommender systems}
\ccsdesc[300]{Mathematics of computing~Probability and statistics}

\keywords{A/B testing, online controlled experiments, two-sided platforms, marketplace interference, recommender systems}

\maketitle
\raggedbottom
\setlength{\emergencystretch}{1em}
\hfuzz=5pt

\section{Introduction}
\label{sec:intro}

Content platforms such as short-video feeds are two-sided marketplaces: creators
supply content and viewers consume it, mediated by a recommender that allocates a
finite pool of viewer attention. A central task on such platforms is
\emph{content exploration}: surfacing new, unconnected content that carries little
or no engagement history, so that promising items and creators can be discovered
rather than starved~\cite{li2010contextual,schein2002coldstart,su2023longterm}. When a platform
changes how it explores (for example, boosting distribution for a targeted set of
creators), the standard way to measure the change is an online controlled
experiment (A/B test). But a naive one-sided A/B test silently violates the
stable-unit-treatment-value assumption (SUTVA): treated and control content
compete for the same viewers, so the control arm is contaminated by the
treatment, and the measured effect diverges from what a full launch would
produce~\cite{johari2022twosided,liu2021budgetsplit}.

The principled remedy is \emph{symmetric two-sided isolation}: randomize both
creators \emph{and} viewers, at a common fraction $p$, into isolated
submarkets, so that the treatment's redistribution of attention plays out
entirely within the treated submarket, removing cross-arm interference by
design. It appears as \emph{user--corpus co-diversion} at
Google~\cite{wang2023fresh}, \emph{budget-split} at
LinkedIn~\cite{liu2021budgetsplit}, \emph{symmetric A/B tests} in
creator-competition studies~\cite{yao2024welfare}, and \emph{two-sided} or
\emph{multiple randomization} more broadly~\cite{johari2022twosided,bajari2021mrd}.
Here \emph{symmetric} means the same sampling fraction on the creator and viewer
sides, and \emph{isolation} means each arm's viewers are served content only from
creators assigned to that arm; matched fractions preserve the supply--demand
ratio but, as we show, not the absolute catalog.

For creator-side treatments such as boosting a creator segment or changing supply incentives, symmetric isolation is therefore the workhorse design (fallback designs fit narrower conditions; \Cref{sec:alternatives}). The design runs at industry scale: teams iterating on such policies open many concurrent symmetric experiments, partitioning the creator and viewer populations into mutually isolated cells, which pushes each cell toward a \emph{small} fraction $p$. Its cost is thus paid on every creator-facing launch decision, and most heavily at the small fractions concurrency forces. 

That cost, however, has been largely overlooked. In
this paper we show that \textbf{merely running a symmetric experiment perturbs
the recommendation ecosystem}, even when there is no treatment at all. In a pure
A/A test, with identical policy in both arms, we observe large, statistically
significant drops in viewer engagement inside the isolated arms: on a production
short-video platform, holding viewer traffic fixed at $10\%$ and shrinking the
creator catalog from $70\%$ to $10\%$ lowers exploration view time by $12.3\%$
and story completions by $19.0\%$; the smaller $2\%$ cell loses $29.8\%$ of view
time. Because there is no treatment, these are not effects; they are
\emph{artifacts of the design itself}. These losses concentrate on
exploration content (whole-feed effects are an order of magnitude smaller),
but exploration metrics are precisely what creator-side and cold-start
experiments are run to measure: the artifact lands on the estimand itself, at
a size that rivals typical treatment effects.

We identify per-viewer catalog thinning as an important contributor: isolation
reduces each viewer's addressable catalog to the in-arm fraction, however
implemented, and a thinner catalog offers fewer good candidates, so the
best-matched items surfaced are worse. We model this \emph{match-quality}
degradation with the order statistics of a viewer's match quality.

Prior work treats two-sided isolation as the remedy for marketplace interference and measures the bias it removes (\Cref{sec:related}); we measure the remedy's own cost, pairing theory with practice: we prove closed-form loss laws \emph{and} calibrate and test them on the A/A sweep above and an independent catalog-ablation experiment. Even at a large catalog scale the effect is pronounced, for an instructive reason: the new, unconnected content relevant to a given viewer is only a thin slice of the
catalog, so filtering it to a few percent of creators leaves too few strong candidates. To our knowledge this is the first production-scale measurement of the ecosystem cost of a symmetric two-sided
experiment.

Our contribution is diagnostic rather than a new serving system: a production-calibrated measurement, on live traffic, of a failure mode of a widely used experimental design, plus a reusable preflight procedure for deciding when the design can be trusted. We study the \emph{content-side} cost: the reduction in measured viewer engagement when the catalog is thinned; these metrics are the readouts whose fidelity is at stake, not an objective the design advances. Creator-side artifacts enter only as an ancillary illustration of the dual (\Cref{sec:empirical}), with creator retention and the causal decomposition of supply responses left to a companion study. Concretely:
\begin{enumerate}[leftmargin=1.5em,itemsep=2pt,topsep=2pt]
  \item \textbf{Model.} An order-statistics model of the
  per-viewer engagement loss caused by symmetric isolation (\Cref{sec:model}).
  \item \textbf{Tail-class laws.} A characterization, via extreme-value theory,
  of three loss regimes, logarithmic (light tails), scale-free (heavy tails), and vanishing (bounded support), with complete proofs (\Cref{prop:exp,thm:tail}).
  \item \textbf{Scaling and regimes.} Traffic floors and a
  match-quality-vs.-supply-limited boundary, characterizing when a larger
  platform makes isolation cheaper (\Cref{cor:scaling}).
  \item \textbf{Experiments and guidance.} Support from two independent
  production experiments (a symmetric A/A traffic sweep and a pure catalog
  ablation consistent with the calibrated model) plus Monte Carlo simulation,
  and a preflight procedure that returns the smallest trustworthy traffic
  fraction, with fallback designs when none exists (\Cref{sec:validation,sec:guidance}).
\end{enumerate}

\section{Related Work}
\label{sec:related}
The closest line of work studies marketplace interference: randomizing a
single side produces bias whose sign and magnitude depend on market
balance~\cite{johari2022twosided}; two-sided
randomization measures and corrects the competition effects that drive the
bias~\cite{li2022interference,johari2024when}. Budget-split
designs isolate shared-budget competition and report interference bias up to
$230\%$ of the effect size~\cite{liu2021budgetsplit}; multiple randomization
designs generalize randomization across interacting
populations~\cite{bajari2021mrd,nandy2021twosided}. Closest to our setting,
Google's fresh-content work~\cite{wang2023fresh} co-diverts users and corpus to
measure fresh/cold-start treatment effects without leakage, arguing that
proportional user/corpus sizing keeps the measured effect consistent with full
deployment. Our symmetric design is a two-sided isolation in this family, but where
prior work uses it to \emph{remove} bias, we study the \emph{cost} the
isolation itself imposes and surface the limit of the proportionality
argument: under heavy-tailed relevance, co-diversion carries a scale-free
engagement cost that consistency arguments miss.

A second line pursues global treatment effects under network interference:
graph cluster randomization isolates clusters~\cite{ugander2013graph}, and meta-experiments
that ``randomize over randomized experiments'' compare designs side by
side~\cite{saveski2017detecting,holtz2025clusterrandomization}; our A/A traffic
sweep is a meta-experiment of this kind, used to isolate the design's own cost.

A third line concerns the supply side: feedback loops contaminate
seller-side tests~\cite{sellerside2024}; in creator-side experiments, treated
and control creators compete for exposure, making costly double-sided randomization the unbiased benchmark~\cite{zhan2024creatorside}. Industrial cold-start systems lean on creator-side experiments for producer-incentive and ecosystem effects~\cite{chen2025kuaishou}, and
ego-cluster designs target creator metrics at a traffic cost that itself
motivates efficiency work~\cite{su2023egocluster}. This line
treats double-sided isolation as the \emph{gold standard} against interference;
our complementary message is that the gold standard itself carries a measurable
content-side cost.

\section{Setting}
\label{sec:setting}
A content platform (e.g.\ a short-video app) hosts a set of $\nactive$ active creators, $\creators$, and a viewer population $\viewers$. For each request in a viewing session, a retrieval stage returns a ranked set of candidates from a shared index, and a ranker surfaces the top items, whether as a slate, a page, or a continuous scroll; we say \emph{feed}. The feed is 
\emph{mixed}:       
exploration content (\Cref{sec:intro}) is interleaved with the viewer's personalized content, so the feed is always filled regardless of exploration-content supply.

A symmetric experiment at fraction $p$ randomly partitions creators and
viewers into two disjoint arms, treatment and control, each holding a fraction $p$ of both sides (remaining units are outside the experiment), and
isolates the resulting submarkets: treated viewers are served only treated
creators' content, and control viewers only control creators' content. \textbf{Isolation
thins the in-arm candidate pool.} Enforcing it at scale costs scanning of a shared index, and whether
candidate supply survives thin fractions depends on where the filter sits in
the retrieval path (\Cref{sec:guidance}). 

To isolate the cost of the design from any treatment effect, we run a pure
\emph{A/A} experiment: every arm applies the identical production policy, and we vary how much catalog and audience each arm retains against a thicker-catalog reference arm (itself isolated). Any systematic difference from that
reference is therefore an artifact of isolation, not a treatment effect. The identifying contrast deliberately unties what the symmetric design ties together, varying the catalog share at a fixed viewer share; a smaller symmetric cell serves as a separate stress test.

\section{An Ecosystem-Cost Model}
\label{sec:model}

We model the engagement loss induced by symmetric isolation from first
principles: isolation thins the candidate pool, and we quantify the resulting
loss of best-match quality as a function of the isolation fraction, the
platform size, and, crucially, the \emph{tail} of match quality.

\subsection{Model and assumptions}
\label{sec:model-setup}

\begin{definition}[Match quality and selection]
\label{def:model}
For a fixed viewer, each candidate item $i$ in the full addressable pool, of
size $\poolsize$ (the \emph{platform size}), carries a \emph{match quality}
$\util_i$ (its realized relevance and engagement potential for that viewer),
modeled as i.i.d.\ draws $\util_1,\dots,\util_{\poolsize}\sim F$
from the catalog's match-quality distribution $F$. The recommender surfaces the
highest-quality available item, so per-request engagement is a non-decreasing
function $g(\cdot)$ of the top order statistic
$\ordstat{m}\eqdef\max_{i\le m}\util_i$ over the $m$ addressable candidates. We
take $g=\mathrm{id}$ for the leading analysis and treat slates in \Cref{sec:slate}.
\end{definition}

Throughout, $\util$ is anchored to a positive \emph{observable} engagement
scale (e.g.\ expected view time contributed), not to a latent score defined
only up to monotone transformation, for which the ratio in \Cref{eq:loss}
below would be meaningless. Because engagement metrics are bounded, a ``heavy
tail'' is an effective description of $F$'s upper quantiles over the pool sizes
an experiment probes; a binding cap calls for a finite-endpoint model, of which
\Cref{thm:tail} treats the standard polynomial-endpoint (Weibull) case.

Isolation enters the model through a single channel: it does not alter $F$ or
the viewer's tastes, it only reduces the number of candidates $m$ from which
the best match is drawn. Because creators are retained in blocks, the in-arm
pool size is random; it nevertheless concentrates:

\begin{lemma}[Thinning concentration]
\label{lem:thin}
Under symmetric isolation at fraction $p$ (each creator's block of items
retained independently with probability $p$), the in-arm pool size $S$ is
random, with best in-arm match $M_S=\max_{i\le S}\util_i$. Suppose the largest
creator's share of the catalog vanishes as $\poolsize\to\infty$, and the
expected best match scales predictably with pool size:
$\E[\ordstat{cm}]/\E[\ordstat{m}]\to c^{\rho}$ for some $\rho\ge0$
($\rho=1/\alpha$ for the heavy tail below, $\rho=0$ otherwise). Then, at
fixed $p\in(0,1]$, $\E[M_S]=\E[\ordstat{p\poolsize}]\,(1+o(1))$: the random
in-arm pool may be replaced by a deterministic pool of exactly $p\poolsize$
items.
\end{lemma}
\noindent All proofs are in Appendix~\ref{app:proofs}. Define the \emph{engagement loss}
\begin{equation}
\loss(p) \;\eqdef\; 1 - \frac{\E[\ordstat{p\poolsize}]}{\E[\ordstat{\poolsize}]}.
\label{eq:loss}
\end{equation}
$\loss(p)$ is the relative reduction in expected best-match quality caused by
thinning. By \Cref{lem:thin} it depends only on the effective pool size $p\poolsize$; assuming the serving path effectively selects from the full
in-arm pool, the results below are implementation-agnostic.

\subsection{Loss law is set by the tail of match quality}
\label{sec:tail-law}

Write $\phi(m)=\E[\ordstat{m}]$ for the expected best match among $m$ candidates.
How fast $\phi$ grows with $m$ is governed by classical extreme-value
theory~\cite{dehaan2006evt}. $F$ itself can be any distribution; just as
normalized \emph{sums} converge to the one Gaussian limit, whenever the
\emph{maximum} of $m$ draws, centered and scaled by sequences $b_m$ and $a_m$
(``norming constants''), converges to a non-degenerate limit, that limit is
one of exactly three laws: Fr\'echet, Gumbel, or reversed Weibull. Which one applies depends only on the
upper tail of $F$, measured by the tail function $\bar F(x)\eqdef1-F(x)$, the
probability that a single candidate exceeds quality $x$; this sorts
distributions into three families (``max-domains of attraction''), each named
after its limit law. Under mild moment conditions
$\phi(m)=b_m+a_m\mu+o(a_m)$ with $\mu$ the limit's mean, a fixed constant
($\gamma\approx0.577$ for the Gumbel family, $\Gamma(1-1/\alpha)$ for
Fr\'echet, $-\Gamma(1+1/\alpha)$ for the bounded case). Substituting into
\Cref{eq:loss} yields the loss laws below: two Gumbel-domain benchmarks
(exponential and lognormal), sharp results for the Fr\'echet and Weibull
domains, and a general Gumbel-domain formulation (\Cref{rem:gumbel}).

\begin{proposition}[Gumbel-domain benchmarks: exponential and lognormal]
\label{prop:exp}
(i) If $F=\mathrm{Exp}(\lambda)$, then $\phi(m)=\lambda^{-1}(\ln m+\gamma)+O(1/m)$ and
\begin{equation}
\loss(p)=\frac{\ln(1/p)}{\ln\poolsize+\gamma}
+O\!\Big(\tfrac{1}{\poolsize\ln\poolsize}\Big),
\label{eq:gumbel}
\end{equation}
logarithmic in $1/p$ and vanishing slowly ($\propto 1/\ln\poolsize$) with
$\poolsize$. (ii) If $F$ is lognormal, $\ln\util\sim\mathcal{N}(\nu,\sigma^2)$
(a common engagement-duration model~\cite{yin2013dwell}), then at fixed $p$
\begin{equation}
\loss(p)=\frac{\sigma\,\ln(1/p)}{\sqrt{2\ln\poolsize}}\,(1+o(1)),
\label{eq:lognormal}
\end{equation}
the same logarithmic shape in $1/p$ with still slower relief
($\propto1/\sqrt{\ln\poolsize}$).
\end{proposition}

\begin{theorem}[Regularly varying tails]
\label{thm:tail}
Fix $p\in(0,1)$ and let $\poolsize\to\infty$.
\begin{enumerate}[nosep,leftmargin=*]
  \item \textbf{Heavy tail (Fr\'echet domain).} If
  $\bar F(x)=x^{-\alpha}\ell(x)$ with $\alpha>1$ and $\ell$ slowly varying,
  $\ell(cx)/\ell(x)\to1$ (e.g.\ Pareto, Zipf, and Student-$t$ distributions),
  then $\phi$ is regularly varying of index $1/\alpha$
  ($\phi(cm)/\phi(m)\to c^{1/\alpha}$, as in \Cref{lem:thin}), and
  \begin{equation}
  \loss(p)=1-p^{1/\alpha}(1+o(1)),
  \label{eq:frechet}
  \end{equation}
  a power law \emph{independent of platform size} $\poolsize$, depending on
  $F$ only through the tail index $\alpha$ (the slowly varying $\ell$ cancels
  in the loss ratio).
  \item \textbf{Bounded support (Weibull domain).} If $F$ has a finite right
  endpoint $u^\ast$ with $\bar F(u^\ast-t)\sim c\,t^{\alpha}$ (e.g.\ the Beta
  distribution), then $u^\ast-\phi(m)$ is regularly varying of index
  $-1/\alpha$, $\phi(m)=u^\ast-c'm^{-1/\alpha}(1+o(1))$, with $c'=\Gamma(1+1/\alpha)\,c^{-1/\alpha}$, and

  \begin{equation}
  \loss(p)=\frac{c'}{u^\ast}\poolsize^{-1/\alpha}\bigl(p^{-1/\alpha}-1\bigr)(1+o(1)),
  \label{eq:weibull}
  \end{equation}
  vanishing polynomially with platform size.
\end{enumerate}
\end{theorem}

\begin{remark}[The third family: the Gumbel domain]
\label{rem:gumbel}
For a general $F$ in the Gumbel domain with infinite right endpoint (e.g.\ the
exponential, Gaussian, and
lognormal distributions), given moment convergence of the normed maxima
(Appendix~\ref{app:proofs}), \Cref{eq:loss} gives
$\loss(p)=\bigl(1-b_{p\poolsize}/b_{\poolsize}\bigr)(1+o(1))$ at fixed $p$,
whose shape depends on the norming sequence $b_m$. The constant of the
logarithmic law~\eqref{eq:gumbel} is specific to $b_m\sim c\ln m$
(exponential-type tails), but the logarithmic order is more generic: Gaussian
match quality obeys it with about half the constant of \eqref{eq:gumbel}, and
the lognormal norming recovers \Cref{eq:lognormal} (derivations in
Appendix~\ref{app:proofs}). Within one family, platform-size relief thus
ranges from order $1/\ln\poolsize$ to the more slowly vanishing
$1/\sqrt{\ln\poolsize}$, in the lognormal case so
slow that platform growth buys little in practice. Power laws and
lognormals are famously hard to distinguish in
data~\cite{mitzenmacher2004}; the practical agreement
of \Cref{eq:frechet,eq:lognormal} means the message survives the ambiguity.
\end{remark}

\Cref{prop:exp,thm:tail} carry a sharp practical message: \textbf{whether a
larger platform reduces the cost of isolation depends on the tail of match
quality.} For the Gumbel-domain and bounded families
(\Cref{eq:gumbel,eq:lognormal,eq:weibull}) the loss shrinks with $\poolsize$;
but for
heavy-tailed match quality (\Cref{eq:frechet}) the loss is \emph{scale-free},
and platform size gives no relief.

Heavy tails are an empirically relevant regime. A recommender maximizes over a
large candidate pool, and popularity skew (classically power-law in
user-generated video~\cite{cha2007tube}), semantic affinity, and retrieval
diversity make a heavy upper tail in the winner's per-viewer utility
plausible; engagement itself is skewed enough that
watch-time predictors model quantiles rather than
means~\cite{zhan2022duration}.

\subsection{Slates and top-\texorpdfstring{$k$}{k} selection}
\label{sec:slate}

Whether served as a slate or consumed sequentially in a scroll session,
engagement aggregates the top $k$ order statistics, which share the maximum's
norming constants; the asymptotics extend term-by-term. We treat $k$ and any
position weights as fixed (weights are absorbed into constants); random
session length or stopping rules are outside the model.

\begin{proposition}[Slate invariance]
\label{prop:slate}
For fixed $k$, let
$\loss_k(p)\eqdef1-\E\bigl[\sum_{j\le k}\util^{(j)}_{(p\poolsize)}\bigr]
/\E\bigl[\sum_{j\le k}\util^{(j)}_{(\poolsize)}\bigr]$, where
$\util^{(j)}_{(m)}$ is the $j$-th largest of $m$ draws. In each family above,
$\loss_k(p)$ obeys the same functional form as $\loss(p)$, with
tail-class-specific constants; the scaling in $p$ and $\poolsize$ is unchanged.
\end{proposition}

\begin{remark}[Creator-side dual]
\label{rem:dual}
Isolation thins an item's addressable \emph{audience} exactly as it thins a
viewer's \emph{catalog} ($\util$ is a property of the pair). A viewer's
engagement is driven by a \emph{maximum} over items, whereas an item's
received engagement sums over the viewers whose requests it wins; yet in a
matched-fraction cell the \emph{mean} received engagement per creator falls
by the same $\loss(p)$, with no allocation model needed
(Appendix~\ref{app:dual}). How the loss distributes \emph{across} creators
does need one, so beyond the mean we use the dual only directionally.
\end{remark}

\subsection{Traffic floors and platform scaling}
\label{sec:scaling}

Two ecosystem-quality constraints bound the traffic a trustworthy symmetric
experiment needs, and statistical power adds a third (\Cref{sec:power}). Each is a \emph{floor}: the smallest isolation fraction $p$ satisfying the constraint. 

\textbf{Supply floor.} The order-statistics model needs a non-degenerate pool: a request is well served
only if the in-arm pool holds at least one candidate above an acceptable
match-quality bar $q_0$. In a pool of $m$ candidates,
$\Prob(\text{none above } q_0)=(1-\bar F(q_0))^m\le e^{-m\bar F(q_0)}$, so
guaranteeing an above-bar candidate with probability $\ge1-\delta$ gives the
sufficient (mildly conservative) \emph{viable-supply threshold}
\begin{equation}
\suppmin \;=\; \frac{\log(1/\delta)}{\bar F(q_0)},
\label{eq:mstar}
\end{equation}
small when good content is common, large when it is rare. Writing the pool as
$\poolsize=\kappa\nactive$ (with $\kappa$ addressable candidates per active
creator before the quality bar, assumed stable as $\nactive$
grows), the requirement
$p\poolsize\ge\suppmin$ becomes the check
\begin{equation}
p\,\nactive \;\ge\; \frac{\suppmin}{\kappa}\;=:\;\tilde m,
\qquad\text{i.e.}\qquad
\pmin_{\text{supply}} \;=\; \frac{\tilde m}{\nactive} \;\propto\; \frac{1}{\nactive},
\label{eq:supplyfloor}
\end{equation}
enough in-arm creators to clear the quality bar. The floor's $1/\nactive$
scaling is tail-class-independent, though its constant depends on $F$, $q_0$,
and $\delta$; below it the above-bar guarantee fails and scarcity and backfill
mechanisms may dominate (\Cref{sec:regimes}).

\textbf{Tolerance floor.} For tolerance $\varepsilon$ in the match-quality regime,
\Cref{prop:exp,thm:tail} imply:
\begin{equation}
\pmin_{\varepsilon} \;=\;
\begin{cases}
e^{-\gamma\varepsilon}\,\poolsize^{-\varepsilon}, & \text{exponential,}\\[2pt]
(1-\varepsilon)^{\sqrt{2\ln\poolsize}/\sigma}, & \text{lognormal,}\\[2pt]
(1-\varepsilon)^{\alpha}, & \text{heavy (Fr\'echet),}\\[2pt]
\bigl(1+\tfrac{\varepsilon u^\ast}{c'}\poolsize^{1/\alpha}\bigr)^{-\alpha}, & \text{bounded (Weibull).}
\end{cases}
\label{eq:tolfloor}
\end{equation}
These floors are asymptotic, valid while $\pmin_\varepsilon\poolsize\gg1$
(automatic except in the bounded case); in the bounded case
$\pmin_\varepsilon\poolsize=O(1)$, so the constant is a small-$\varepsilon$
approximation and the exact floor follows by inverting $\phi$. The lognormal
row is exact for $\ln\pmin_\varepsilon$ to leading order (it omits a constant
factor $e^{d_\varepsilon^2/2}$, $d_\varepsilon=-\ln(1-\varepsilon)/\sigma$),
and the exponential row's constant is specific to $F=\mathrm{Exp}$
(derivations in Appendix~\ref{app:floors}).

\begin{corollary}[When platform size helps]
\label{cor:scaling}
The binding floor across the two ecosystem-quality
constraints is $\pmin=\max\{\pmin_{\text{supply}},\pmin_\varepsilon\}$,
and how it falls with platform size depends on the tail. For bounded tails both
floors decay as $1/\nactive$, so $\pmin\propto1/\nactive$. For exponential-type
tails the tolerance floor decays only as $\nactive^{-\varepsilon}$ (up to
slowly varying factors), slower
than the $1/\nactive$ supply floor, so it binds at scale and
$\pmin=\nactive^{-\varepsilon+o(1)}$; for the lognormal it decays more slowly
still, slower than any power of $\nactive$:
$(1-\varepsilon)^{\sqrt{2\ln\nactive}/\sigma}$ up to constants. For heavy tails $\pmin_\varepsilon$ is
constant in $\nactive$: the match-quality cost never vanishes, and only the
supply floor improves with scale.
\end{corollary}
\noindent Because two disjoint arms must fit, a symmetric design is feasible only if $\pmin\le\tfrac12$; if $\pmin>\tfrac12$, no allocation satisfies both ecosystem-quality constraints.

\subsection{Two regimes and the line between them}
\label{sec:regimes}

\Cref{eq:supplyfloor} delineates the operating regimes: in the
\textbf{match-quality regime} ($p\poolsize\ge\suppmin$) the pool is thick and
\Cref{thm:tail} applies with the tail-appropriate law; in the
\textbf{supply-limited regime} ($p\poolsize<\suppmin$, or within niche categories
where content is intrinsically scarce) the above-bar guarantee fails.
Operationally we then expect backfill with off-preference content and a binding
cross-viewer supply channel, mechanisms outside the order-statistics model
under which loss should grow faster than the tail laws predict; because
scarcity is category-specific, check
the boundary per interest cluster.

\subsection{Statistical power and variance}
\label{sec:power}
Isolation also drives power: each arm draws on $\approx p\,|\viewers|$ viewers.
\begin{proposition}[Variance and MDE scaling]
\label{prop:power}
For a viewer-level metric with per-viewer variance $\sigma^2(p)$ in each arm
and negligible within-arm cross-viewer covariance, the
treatment-effect estimator has
$\mathrm{Var}\approx 2\sigma^2(p)/(p\,|\viewers|)$, so the minimum
detectable effect (MDE) satisfies
\begin{equation}
\mathrm{MDE}(p)\;\propto\;\frac{\sigma(p)}{\sqrt{p\,|\viewers|}}.
\label{eq:mde}
\end{equation}
\end{proposition}
\noindent Unlike the level shift, which is common to both arms and cancels to
first order (\Cref{sec:discussion}), the $1/\sqrt{p}$ factor does not cancel; power is therefore a first-order reason to raise $p$ and enters the sizing rule of \Cref{sec:guidance}.

\section{Experiments}
\label{sec:validation}
The model of \Cref{sec:model} makes three testable predictions: engagement
falls with the retained catalog share alone and, as a mechanism signature,
falls hardest on the metrics that track best-match quality; the loss is strongly
sub-proportional in the catalog drop, following a tail-class law of
\Cref{sec:tail-law}; and a law calibrated under one design predicts outcomes
under another. Per-viewer match quality is unobserved in our experiments,
so the tail class cannot be read off directly (\Cref{rem:gumbel}; tail diagnostics in
Appendix~\ref{app:tails}); we calibrate the heavy-tail law, the conservative
benchmark whose cost does not vanish with scale, and the only law that
transfers across pools without knowing their size, and test its predictions
out of sample. The choice costs little even if wrong: under the lognormal
alternative of \Cref{eq:lognormal}, a hundredfold growth of the candidate pool
multiplies the loss by only about $0.9$ at the pool sizes our experiments probe
(a $19\%$ completion loss becomes $17\%$); over any attainable growth, the tail
classes we cannot distinguish prescribe the same budget. On a large production
content platform, the A/A sweep of \Cref{sec:empirical} tests the thinning
prediction and calibrates the law; the catalog ablation of
\Cref{sec:ablation} tests sub-proportionality and transfer; simulation
(\Cref{sec:simulation}) verifies the tail-law and traffic-floor calculations.

\subsection{Production A/A: the cost of catalog thinning}
\label{sec:empirical}
We run a pure A/A (identical production policy in every arm). The arms
differ only in how much catalog and audience each isolates:
\begin{itemize}[nosep,leftmargin=*]
  \item \textbf{Reference:} 10\% of viewers served from 70\% of creators (a thick
  catalog);
  \item \textbf{Thin-catalog arm:} 10\% of viewers served from 10\% of creators;
  \item \textbf{Small cell:} 2\% of viewers served from 2\% of creators.
\end{itemize}
The three arms above form the engagement study; a separate, more recent
one-week repeat at $5$--$30\%$ creator fractions supplies only the
serving-path diagnostics of \Cref{tab:retrieval}.
Arms are mutually exclusive random slices of the same viewer and creator
populations; remaining traffic serves the production default, and
viewer-side metrics are per-user means. The arms thin the \emph{exploration content} (new or
under-explored posts collecting early engagement signals), so we report its
engagement.
Across the two-month run, the exploration pool spans a large and continuously
refreshed set of creators and candidate items; \Cref{tab:sweep}
reports a one-week late-run window; losses were stable throughout the run,
arguing against a transient novelty effect.

We use the fixed-viewer contrast for identification, reserve the $2\%$ cell
as a lower-fraction probe, and let simulation supply dense variation; the
marketplace cost of denser sweeps is discussed in \Cref{sec:discussion}.

\begin{figure}[t]
\centering
\includegraphics[width=0.8\columnwidth]{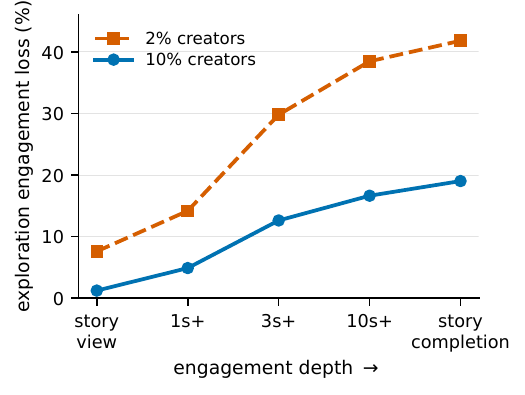}
\caption{Identification via depth. Holding the viewer sample fixed at 10\% and
thinning the catalog (10\% vs.\ 70\% creators), exploration-content engagement loss
\emph{deepens monotonically with engagement depth}: shallow views barely move,
while completion loss is largest. The gradient is consistent with a thinner best-match pool
(\Cref{sec:tail-law}); the further-thinned 2\% cell shows the same pattern, larger.}
\Description{Bar chart of engagement loss by metric depth for two catalog fractions: losses grow from shallow views to completions, and are larger in the 2 percent cell.}  
\label{fig:empirical}
\end{figure}

\textbf{Identification.} The thin-catalog arm and the reference each serve an equally sized, independently
randomized 10\% viewer sample and differ only in the in-arm catalog (10\% vs.\
70\% of creators), so engagement differences identify the catalog contrast
rather than a treatment or viewer-sample-size difference.
\Cref{tab:sweep,fig:empirical} show that exploration-content engagement
falls under the thin catalog with a loss that \emph{deepens monotonically
with engagement depth}: shallow story views fall $1.2\%$, one-second views
$4.9\%$, three-second $12.6\%$, ten-second $16.7\%$, and story completions
$19.0\%$. That gradient is consistent with
the order-statistics mechanism (a thinner pool lowers the \emph{best} available
match), though funnel and substitution effects could also produce an increasing
gradient. Several design checks support the causal reading. First, a retrospective
balance check on the week before launch finds the arms well balanced: pre-period
differences on the metrics of \Cref{tab:sweep} are within about $1\%$, an order
of magnitude below the treatment-period losses of $12$--$19\%$ on the deep
metrics. Second, engagement
on established (non-exploration) content, which is not thinned, \emph{rises}
rather than falls (story view time $+0.4\%$, story completions $+0.5\%$), and the gain itself grows with depth, indicating substitution toward
established content rather than a uniform degradation.

\begin{table}[t]
\centering
\caption{Exploration-content engagement under a thinned catalog, relative to the
thick-catalog reference (10\% viewers, 70\% creators). The 10\%-creator arm holds
the viewer sample fixed at 10\%, isolating catalog thinning; the 2\% cell thins
further (and also shrinks the sample). Contrasts are unadjusted
differences in per-user means over a one-week cumulative analysis window,
under viewer-level randomization; $p$-values are per-metric, and every
contrast is significant at $p<10^{-4}$. Arms are large samples of the viewer
population and show no evidence of sample-ratio mismatch~\cite{fabijan2019srm}
($p_{\mathrm{SRM}}>0.6$).}
\label{tab:sweep}
\begin{tabular}{lrr}
\toprule
Exploration metric & 10\% creators (fixed viewers) & 2\% creators \\
\midrule
Story views       & $-1.24\%$  & $-7.61\%$ \\
View time         & $-12.27\%$ & $-29.76\%$ \\
$10^{+}$s views   & $-16.65\%$ & $-38.48\%$ \\
Story completion  & $-19.03\%$ & $-41.82\%$ \\
\midrule
Favorite          & $-18.20\%$ & $-38.24\%$ \\
Send              & $-16.92\%$ & $-38.93\%$ \\
\bottomrule
\end{tabular}
\end{table}

\textbf{Not a ranker-input-count artifact.} Thinning could starve retrieval: if fewer candidates reach ranking,
engagement would fall for reasons unrelated to match quality. Serving logs
from the separate repeat argue against ranker-input starvation as the
primary explanation (\Cref{tab:retrieval}): the isolation filter removes
candidates in proportion to how thin the arm is, yet backfill keeps the
ranker-input slate approximately stable (within $\sim\!4\%$ across cells).
The much larger, depth-graded engagement loss in the primary study is
consequently unlikely to be explained by ranker-input-count starvation alone.

\begin{table}[tb]
\centering
\caption{In-arm candidate supply per request, by creator fraction (separate
one-week serving-path repeat), each row normalized to its own $30\%$-cell value
(absolute counts withheld). The isolation filter removes candidates in
proportion to how thin the arm is, but backfill keeps the slate that actually
reaches ranking approximately stable.}
\label{tab:retrieval}
{\setlength{\tabcolsep}{4pt}
\begin{tabular}{lrrrr}
\toprule
Mean candidates per request (relative) & 5\% & 10\% & 20\% & 30\% \\
\midrule
Removed by isolation filter & $5.8$ & $4.3$ & $3.0$ & $1.0$ \\
Scored by ranker & $0.96$ & $0.99$ & $0.99$ & $1.00$ \\
\bottomrule
\end{tabular}}
\end{table}

\textbf{Calibration.} \Cref{sec:model-setup} anchors $\util$ to an observable engagement scale, so an
arm ratio of per-user engagement is an aggregate proxy for \Cref{eq:loss} rather
than a per-request measurement of it. We invert the conservative heavy-tail benchmark $1-p^{1/\alpha}$ (\Cref{eq:frechet}). Because the losses are
measured against the $70\%$-creator reference rather than the full catalog, the
retained fraction is $p_{\mathrm{rel}}=0.10/0.70\approx0.14$, not $0.10$;
inverting at that fraction gives per-metric indices of $\alpha\approx9$--$15$ on
the deep metrics and $156$ on story views, which stand apart here as they do
throughout; \Cref{tab:transfer} lists the ones we carry forward. The $2\%$ cell serves as a
lower-fraction check and is not used in the fit
because it also changes the viewer fraction. A single contrast cannot identify the tail
class (any one-parameter law fits one point), so we treat this as an
effective-tail \emph{calibration} over the probed pool sizes and test it out of
sample in \Cref{sec:ablation}.

\textbf{Extreme-thinning stress test.} The $2\%$ cell deepens every loss in \Cref{tab:sweep}. At
$p_{\mathrm{rel}}=0.02/0.70\approx1/35$ the calibrated $1-p^{1/\alpha}$ predicts
deep-metric losses of $21.3\%$ for view time and $32.0\%$ for completion against
observed $29.8\%$ and $41.8\%$, gaps of $8$--$10$ points that are far outside
sampling error at these arm sizes. The law therefore under-predicts at this
fraction, though a proportional cost would predict $97\%$. A pool
approaching the supply floor would produce a shortfall in exactly this direction
(\Cref{sec:regimes} and Appendix~\ref{app:steepening}): slates stay full but backfill draws
off-preference content, so the mechanism predicts no deficit in candidate counts,
which is what \Cref{tab:retrieval} finds over its $5$--$30\%$ range. The cell
also shrinks the viewer sample
(\Cref{prop:power}). We read it as a stress test the law largely but not fully
explains, not as an identified regime transition.

\textbf{Creator side.} In a creator-randomized mirror, creators receive
$32.1\%$ less exploration view time in the $10\%\times10\%$ cell and
$46.3\%$ less in the $2\%\times2\%$ cell, relative to a $10\%$-creator,
$70\%$-audience reference. Between the two matched cells, mean received view
time is $20.9\%$ lower at $2\%$, closely mirroring the $19.9\%$ viewer-side
decline implied by \Cref{tab:sweep}, consistent with the conservation dual
(\Cref{rem:dual}); submissions decline $2.4\%$ and $3.6\%$ ($p<0.01$), a
behavioral response the dual does not imply.

\subsection{Catalog ablation: isolating the per-viewer channel}
\label{sec:ablation}

To separate thinning from the isolation machinery, we run an independent
\emph{catalog ablation}. Viewers are randomized into equal-sized arms; at dose $x$, a
deterministic hash of $(v,i)$ removes a viewer-specific fraction $\approx x$ of the
catalog, and retrieval backfills with the next-ranked candidates.
Served count is held constant and every item remains available to $\approx 1-x$ of
viewers, so aggregate supply and creator assignment are unchanged: the design 
targets the per-viewer channel of \Cref{def:model} while holding the cross-viewer
supply channel (\Cref{sec:regimes}) fixed.

We cap the dose at $\sim\!10\%$ to keep scanning, backfill, and serving latency
within normal operating limits; the symmetric study already probes the deeper
$86$--$97\%$ catalog reductions.

\begin{table}[t]
\centering
\caption{Catalog-ablation experiment (viewer-randomized, equal arms): random
per-viewer catalog drops at two doses.
Entries are \%~change in whole-feed per-user means vs.\ control, $\pm$ the
half-width of an unadjusted per-metric normal-approximation 95\% interval; $^{*}$ marks an estimate whose
interval excludes zero. Measured on the whole feed, since the ablation treats the
entire catalog.}
\label{tab:ablation}
\begin{tabular}{lrr}
\toprule
Metric & $-5\%$ & $-10\%$ \\
\midrule
View time        & $-0.29\pm0.71$ & $-1.18^{*}\pm0.71$ \\
Story completion & $-0.01\pm0.74$ & $-0.86^{*}\pm0.73$ \\
Send             & $-0.79\pm1.11$ & $-1.36^{*}\pm1.10$ \\
Story views      & $+0.02\pm0.73$ & $-0.80^{*}\pm0.73$ \\
\bottomrule
\end{tabular}
\end{table}

At the $10\%$ dose every metric in \Cref{tab:ablation} declines with a
$95\%$ interval excluding zero, view time by $1.18\%$; at the $5\%$ dose none
does. Favorite is measured but too noisy to interpret at this exposure, so we
omit it. The response is strongly \emph{sub-proportional}: removing $10\%$ of the catalog lowers
measured engagement by about $1\%$, an order of magnitude below the proportional benchmark,
as the damped order-statistics response implies. Because it uses none
of the isolation machinery, the ablation provides corroborating evidence for
per-viewer thinning and indicates the loss's direction and order of magnitude,
rather than estimating the isolation cost itself.

\textbf{Out-of-sample parameter transfer.} Carried to the ablation via $1-p^{1/\alpha}$ without refitting, the indices
calibrated on the symmetric fixed-viewer comparison predict $10\%$-dose losses for
view time, completion and send that all fall inside the observed $95\%$ intervals
(\Cref{tab:transfer}). That is a calibration from one 
design predicting the outcome of another. The intervals are wide
relative to the effects, so the test separates the predicted loss from no loss but
not one tail index from another; we read it as \emph{consistent with} the
calibrated law, not as a validation of it. Two limits of scope: shallow view counts
respond to delivery volume rather than best-match quality, which their sweep index
of $156$ already signals, so we do not carry it forward; and at the $5\%$ dose the
predicted effects ($0.3$--$0.6\%$) sit inside the measurement resolution, so that
dose neither supports nor contradicts the law.
Finally, the transfer spans two very different pool sizes: the sweep carves the
exploration pool, whereas the ablation treats the full catalog. The heavy-tail law
predicts no attenuation with pool size (\Cref{eq:frechet}) while the light-tail
law does (\Cref{eq:gumbel,cor:scaling}), so the
agreement is consistent with pool-size independence; but one cross-pool
transfer cannot identify the tail class, and tail-shape stability
remains an assumption.

\begin{table}[t]
\centering
\caption{Out-of-sample parameter transfer at the $10\%$ ablation dose. The tail
index $\hat\alpha$ is calibrated on the symmetric study (\Cref{tab:sweep}) and
applied without refitting via $1-p^{1/\hat\alpha}$ at $p=0.90$.
Observations are the whole-feed changes of \Cref{tab:ablation}, $\pm$ the
half-width of a $95\%$ interval.}
\label{tab:transfer}
\begin{tabular}{lrrr}
\toprule
Metric & $\hat\alpha$ & Pred.\ (\%) & Obs.\ (\%) \\
\midrule
View time         & $14.9$ & $-0.71$ & $-1.18\pm0.71$ \\
Story completion  & $9.2$  & $-1.14$ & $-0.86\pm0.73$ \\
Send              & $10.5$ & $-1.00$ & $-1.36\pm1.10$ \\
\bottomrule
\end{tabular}
\end{table}

\subsection{Simulation}
\label{sec:simulation}

\begin{figure*}[t]
\centering
\includegraphics[width=0.64\linewidth]{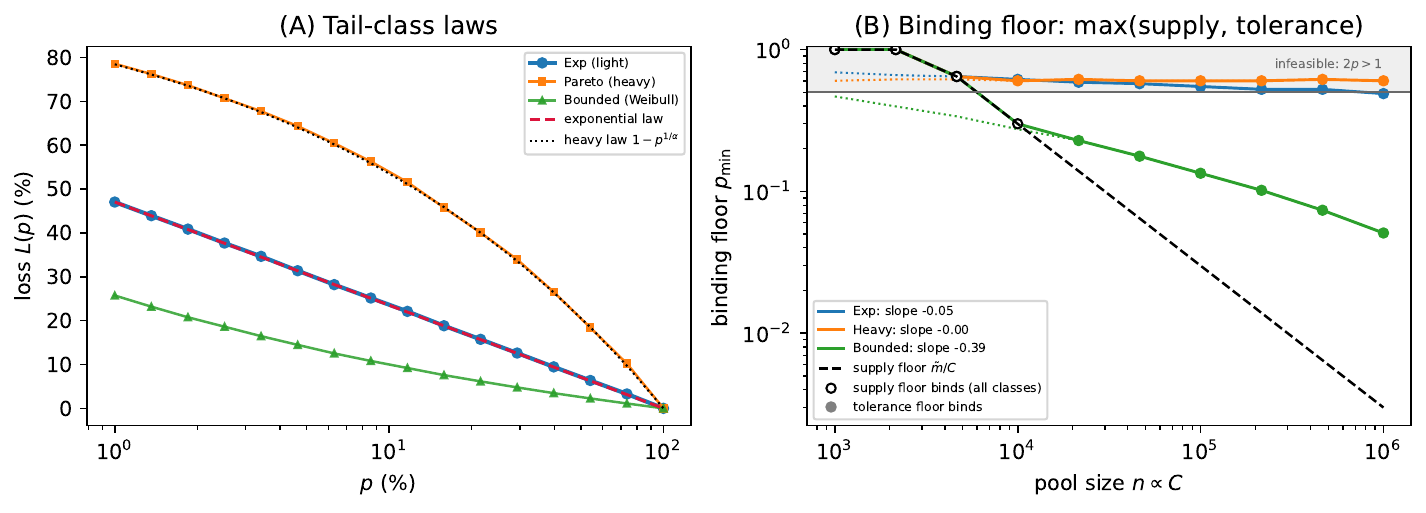}
\caption{Simulation of the order-statistics model. \textbf{(A)} Exponential and Pareto losses at $n=10^4$ match their closed-form laws (dashed: \Cref{eq:gumbel}; dotted: \Cref{eq:frechet}). \textbf{(B)} The binding floor $\pmin=\max\{\pmin_{\text{supply}},\pmin_\varepsilon\}$ (\Cref{cor:scaling}) at $\varepsilon=5\%$ and an illustrative $\suppmin\approx3{,}000$: each class is pinned to the supply floor (dashed) at small $n$, then peels off onto its tolerance floor (dotted) in the corollary's order (open markers: supply floor binds; filled: tolerance floor). The gray band marks two-arm infeasibility ($2p>1$, \Cref{sec:scaling}); the bounded floor is pre-asymptotic here.}   
\Description{Two simulation panels: loss versus isolation fraction matching
closed forms; binding traffic floor versus pool size, flat for the heavy
tail, drifting slowly for the exponential, falling for the bounded class,
with a supply-floor line binding all classes at small pool sizes and a gray region above one half marking two-arm infeasibility.}
\label{fig:simulation}
\end{figure*}

The simulation closes the loop on the theory: panel (A) checks the loss laws
of \Cref{sec:tail-law}, and panel (B) assembles the binding floor of
\Cref{sec:scaling}, the quantity the preflight procedure of
\Cref{sec:guidance} sizes against.
Match utilities are drawn i.i.d.\ from
exponential, heavy-tailed (Pareto: $\alpha=3$ in
panel A; panel B uses $\alpha=10$, within the range calibrated in
\Cref{sec:empirical}), and bounded (Weibull-domain) distributions; we estimate
$\E[\ordstat{m}]$ via the order-statistic identity
$\ordstat{m}\stackrel{d}{=}F^{-1}(B)$, $B\sim\mathrm{Beta}(m,1)$.
\Cref{fig:simulation} shows the results.

\textbf{Findings.} (i) The exponential and Pareto simulations match their
laws to within Monte Carlo error across the full range of $p$; the bounded simulation has the predicted shape but remains
visibly pre-asymptotic at $n=10^4$. The simulated lognormal, qualitatively
similar in shape to the heavy curve over this range, is shown in
Appendix~\ref{app:steepening}.
(ii) The binding floor reproduces \Cref{cor:scaling} operationally, with
two-arm feasibility explicit ($2p\le1$, \Cref{sec:scaling}; gray
band): supply alone requires $n\ge2\suppmin$, and as $n$ grows the classes
part ways. The bounded floor keeps falling (slope $-0.39$) and becomes
feasible once $n\gtrsim2\suppmin\approx6{,}000$ (the first feasible plotted
point is $10^{4}$); the exponential floor drifts at
$n^{-\varepsilon}$ (slope $-0.05$) and reaches the feasibility bound only
near $n=10^{6}$; the heavy-tail tolerance floor is flat (slope $0.00$), so once the supply floor ceases to bind, the binding floor stays at the calibrated ($\alpha=10$) value $p\approx0.6>\tfrac12$: no feasible
symmetric experiment meets a $5\%$ tolerance at any platform size. At looser  tolerances the tolerance floor drops below $\tfrac12$ but remains size-independent and still binds at scale; since concurrent experiment capacity scales as $1/(2p)$ (\Cref{sec:guidance}), it is then a
permanent throughput ceiling, the practical content of the scale-free law.
\section{Choosing a Design in Practice}
\label{sec:guidance}\label{sec:alternatives}

These production findings now inform the planning and traffic sizing of subsequent symmetric experiments. A team predeclares metric-specific loss tolerances, a target MDE, and a candidate-count guardrail, then runs a short calibration sweep, globally or per category. The preflight procedure below estimates the cost, returns the smallest traffic fraction that clears every check, and names a fallback design when no fraction does.

\begin{enumerate}[nosep,leftmargin=*]
  \item \textbf{Calibrate the loss.} Compare a candidate arm against a
  thick-catalog A/A reference and estimate the loss per key metric
  (\Cref{eq:loss,sec:empirical}). A reference that itself retains only a fraction
  $r<1$ of creators measures losses relative to $r$ rather than to the full
  catalog, so calibrate on the relative retention $p/r$
  (Appendix~\ref{app:refrel}). Inverting the heavy-tail
  benchmark then gives an implied $\alpha$ for interpolating between fractions;
  it does not identify the tail class.
  \item \textbf{Size for tolerance and power.} Let $\pmin_\varepsilon$ be the
  smallest fraction satisfying the predeclared loss tolerances
  (\Cref{eq:tolfloor}), and let $\pmin_{\text{power}}$ be the smallest fraction
  meeting the target MDE (\Cref{eq:mde}). Start from
  $\max\{\pmin_\varepsilon,\pmin_{\text{power}}\}$.
  \item \textbf{Apply a candidate-supply guardrail.} The structural supply floor
  scales as $1/\nactive$ (\Cref{eq:supplyfloor}), but its constant depends on
  $F$, $q_0$, $\delta$, and $\kappa$, and must be calibrated locally. In its
  place, log the pre-backfill in-arm yield and the backfill share, and raise
  $p$ whenever a predeclared lower-percentile guardrail fails; the count
  finally scored is insufficient on its own, since backfill can hold slate
  size while degrading match quality (\Cref{tab:retrieval}). This is an
  operational check, not an estimate of $\suppmin$.
  \item \textbf{Stress-test against scale.} A short sweep cannot distinguish tail
  classes, so evaluate the design under the heavy-tail benchmark, for which the
  cost is scale-free (\Cref{cor:scaling}). Size $p$ against the calibrated loss
  rather than the total catalog size.
\end{enumerate}

Evaluating \Cref{eq:tolfloor} at the period-specific calibrated indices makes the trade-off
concrete. A $5\%$ tolerance needs $47\%$ of each population per arm at the
view-time index ($\hat\alpha=14.9$) and admits no valid design at the completion index ($\hat\alpha=9.2$), since two disjoint arms require $2p\le1$; a $10\%$
tolerance needs $21$--$38\%$. Because an experiment consumes $2p$ of both
populations, concurrency scales as $1/(2p)$: at tolerances tight enough for the law to apply, a team can run only one or two symmetric experiments at a time, directly at odds with iteration velocity (\Cref{sec:intro}). Looser tolerances do buy throughput, but they push $p$ toward the few-percent range where \Cref{sec:empirical} finds the law already
under-predicting.

On implementation: per-arm index copies do not scale to concurrent experimentation, so the standard practice is a filter on a shared index. Apply it early, by restricting candidate sources or treating arm membership as a predicate during the retrieval scan, rather than discarding candidates after retrieval, which pays extra scanning and yields the candidate shortfall where scan budgets bind at thin fractions.

If no $p\le\tfrac12$ clears all three checks, symmetric isolation is the wrong
tool, and which check fails points to the replacement: a binding tolerance
floor means per-viewer thinning itself is the problem, so the fallback must
avoid splitting the catalog; a binding candidate guardrail means the market is
thin in absolute terms, and partitioning it further only makes it thinner; if
only power binds, the catalog is adequate, so extend the run or reduce
variance before changing the design. The fallbacks: \emph{cluster randomization} when the catalog partitions into near-self-contained clusters,
whose within-cluster catalogs stay thick~\cite{ugander2013graph,holtz2025clusterrandomization,su2023egocluster};
\emph{budget-split} when the treatment acts through a shared budget or quota
rather than per-item eligibility~\cite{liu2021budgetsplit}; \emph{switchback} when the marketplace is temporally stationary, isolating along the time axis and leaving the catalog intact~\cite{bojinov2023switchback}; and, as a last resort, a \emph{one-sided} test, whose mechanism-dependent interference bias can rival the effect itself~\cite{liu2021budgetsplit,zhan2024creatorside}, read directionally via the treatment's mechanism.

\section{Discussion}
\label{sec:discussion}
In a symmetric A/B (rather than A/A), both arms are isolated at the same $p$,
so the isolation-induced level shift cancels to first order, provided the
treatment does not materially interact with thinning; the cost bites
instead through (i) \emph{external validity}: the effect is measured inside an
ecosystem whose match quality sits $\loss(p)$ below launch conditions;
(ii) \emph{power}: smaller viewer samples and potential under-retrieval add noise; and (iii) the altered content mix shown to assigned traffic for the duration of the experiment. Under a heavy tail the issue can become feasibility rather than merely cost: if the size-independent tolerance floor $\pmin_\varepsilon=(1-\varepsilon)^\alpha
>\tfrac12$, no two-arm symmetric experiment meets the chosen tolerance at any platform size (\Cref{sec:scaling}).

Our production identification uses one fixed-viewer catalog contrast (10\%
viewers with a 10\% vs.\ 70\% creator catalog) plus a smaller $2\%\times2\%$
symmetric cell. A denser catalog-fraction sweep would sharpen the scaling
estimate, but in a two-sided production marketplace each additional point consumes
scarce creator mass and viewer traffic under a deliberately thinned catalog. Unlike user-side experimentation, where arms mainly
reallocate users, the empirical budget is constrained by the same
finite-supply mechanism that creates the isolation cost; hence we use production data to identify the mechanism and
calibrate magnitude, and simulation for clean-variation scaling.

We close with this study's limitations. The model assumes
i.i.d.\ match utilities and top-item selection; under standard
weak-dependence (extremal-index) conditions, correlated candidates change the
order-statistics constants, while strong dependence could alter the scaling
itself. We study A/A artifacts and reason about treatment experiments only
indirectly, through the implications above. While the production
contrasts calibrate an effective heavy-tail index over the observed range,
they do not identify the max-domain (\Cref{rem:gumbel}), an ambiguity that affects the
asymptote rather than the sizing advice. The serving system itself
evolves, so we read $\hat\alpha$ as a calibration of the operating range and
period rather than a platform constant. We also do not separately identify
serving-side effects of deeper scanning (e.g.\ latency); the
dose-capped ablation and the established-content substitution argue they are
not the dominant channel. Finally, our primary estimand is the
\emph{content-side} cost: the creator-side A/A artifacts
(\Cref{sec:empirical}) illustrate the conservation dual (\Cref{rem:dual}),
but creator retention and the causal decomposition of these outcomes,
mechanical exposure loss versus behavioral supply response, are left to a
companion study.

\section{Conclusion}
\label{sec:conclusion}
Symmetric two-sided isolation removes cross-arm marketplace interference, but even in a pure A/A it thins each viewer's catalog and measurably reduces engagement. Our order-statistics model shows that platform scale relieves this cost for bounded tails, only slowly in the Gumbel domain, and not at all under heavy tails; the resulting traffic floors can make a two-arm design infeasible outright. Two production studies corroborate the mechanism and calibrate its magnitude; simulation verifies the calculations. Practitioners should estimate the cost before launch, size traffic accordingly, and switch designs when no feasible fraction remains.

% \begin{acks}
% TODO: acknowledgments for camera-ready.
% \end{acks}

\message{^^JCONTENT-ENDS-ON-PAGE: \thepage^^J}
\bibliographystyle{ACM-Reference-Format}
\bibliography{references}

\appendix
\section{Proofs}
\label{app:proofs}

Throughout, write $\phi(m)\eqdef\E[\ordstat{m}]=\E[\max_{i\le m}\util_i]$ with
$\util_i\stackrel{\text{iid}}{\sim}F$, and recall $\loss(p)=1-\phi(p\poolsize)/\phi(\poolsize)$.

\subsection{Proof of \texorpdfstring{\Cref{lem:thin}}{Lemma 1}}
Write $S=\sum_c s_c B_c$, where $s_c$ is creator $c$'s item count,
$\sum_c s_c=\poolsize$, and $B_c\stackrel{\text{iid}}{\sim}\mathrm{Bernoulli}(p)$.
Condition on $S$: with the convention $M_0=0$ on the empty-pool event,
$\E[M_S]=\E_S[\phi(S)\,\mathbf{1}\{S\ge1\}]$, and
$\Prob(S=0)=(1-p)^{K}$, with $K$ the number of creators, is exponentially small
and absorbed into the $o(1)$ terms below (non-integer $p\poolsize$ is handled
by rounding, negligible under regular variation). In each tail class of \Cref{thm:tail},
$\phi$ is regularly varying at infinity with index $\rho\ge 0$
($\rho=0$ for the light and bounded tails, $\rho=1/\alpha$ for the heavy tail),
i.e.\ $\phi(cm)/\phi(m)\to c^{\rho}$ for $c>0$. Here
$\E[S]=p\poolsize$ and
$\mathrm{Var}(S)=p(1-p)\sum_c s_c^2\le p(1-p)\,\poolsize\max_c s_c$, so
$\mathrm{Var}(S/(p\poolsize))=O(\max_c s_c/\poolsize)\to0$ under the
vanishing-share condition; Chebyshev gives
$S/(p\poolsize)\to 1$ in probability.
By the uniform convergence theorem for regularly varying
functions~\cite{dehaan2006evt}, $\phi(S)/\phi(p\poolsize)\to 1$ in probability.
The ratio is uniformly bounded: $\phi$ is non-decreasing and $S\le\poolsize$, so
$\phi(S)/\phi(p\poolsize)\le\phi(\poolsize)/\phi(p\poolsize)\to p^{-\rho}$.
Bounded convergence then upgrades convergence in probability to convergence in
mean: $\E_S[\phi(S)\,\mathbf{1}\{S\ge1\}]/\phi(p\poolsize)\to 1$. \qed

\subsection{Proof of \texorpdfstring{\Cref{prop:exp}}{Proposition 1}}
\paragraph{(i) Exponential, exact.}
For i.i.d.\ exponentials, the maximum has the representation
$M_m \stackrel{d}{=}\tfrac{1}{\lambda}\sum_{i=1}^{m}\tfrac{E_i}{i}$ with $E_i$ i.i.d.\
$\mathrm{Exp}(1)$ (R\'enyi), hence
\[
\phi(m)=\E[M_m]=\frac{1}{\lambda}\sum_{i=1}^{m}\frac1i=\frac{H_m}{\lambda}
=\frac{\ln m+\gamma+O(1/m)}{\lambda},
\]
where $H_m$ is the $m$-th harmonic number and $\gamma$ is Euler's constant. Therefore
\[
\loss(p)=1-\frac{\ln(p\poolsize)+\gamma}{\ln\poolsize+\gamma}
=\frac{\ln\poolsize-\ln(p\poolsize)}{\ln\poolsize+\gamma}
=\frac{\ln(1/p)}{\ln\poolsize+\gamma}+O\!\Big(\tfrac{1}{\poolsize\ln\poolsize}\Big),
\]
which is \Cref{eq:gumbel}. The $\gamma$ here is the mean of the standard
Gumbel limit, entering through $\phi(m)=b_m+\gamma a_m+o(a_m)$ with
$b_m=\lambda^{-1}\ln m$ and $a_m=\lambda^{-1}$; the general mechanism is the
$\Pi$-variation argument below.

\paragraph{(ii) Lognormal.}
For $\ln\util\sim\mathcal{N}(\nu,\sigma^2)$, take $b_m=e^{\nu+\sigma z_m}$
with $z_m=\Phi^{-1}(1-1/m)$ the standard normal upper quantile and
$a_m=\sigma b_m/\sqrt{2\ln m}$. Standard moment convergence for lognormal
maxima~\cite{dehaan2006evt} gives
$\phi(m)=b_m\bigl(1+O(1/\sqrt{\ln m})\bigr)$.
From the quantile expansion
$z_m=\sqrt{2\ln m}-\tfrac{\ln\ln m+\ln 4\pi}{2\sqrt{2\ln m}}
+o\bigl((\ln m)^{-1/2}\bigr)$,
\[
z_{p\poolsize}-z_{\poolsize}
=-\frac{\ln(1/p)}{\sqrt{2\ln\poolsize}}\,(1+o(1))
\quad\text{at fixed }p,
\]
and the $\gamma a_m$ corrections change the ratio
$\phi(p\poolsize)/\phi(\poolsize)$ only at order
$\ln(1/p)(\ln\poolsize)^{-3/2}$, which is $o(\loss(p))$. Hence
\[
\loss(p)=1-e^{\sigma(z_{p\poolsize}-z_{\poolsize})}\,(1+o(\loss(p)))
=\frac{\sigma\ln(1/p)}{\sqrt{2\ln\poolsize}}\,(1+o(1)),
\]
which is \Cref{eq:lognormal}. \qed

\paragraph{General Gumbel-domain losses ($\Pi$-variation).}
The two benchmarks, and the Gaussian case of \Cref{rem:gumbel}, are
instances of one mechanism. Write $U(t)\eqdef F^{-1}(1-1/t)$, so $b_m=U(m)$.
$F$ lies in the Gumbel domain iff $U$ is $\Pi$-varying:
$U(tx)-U(t)=a(t)\ln x\,(1+o(1))$ for an auxiliary function
$a(\cdot)$~\cite{dehaan2006evt}. When the right endpoint is infinite,
$U(t)\to\infty$, $a(t)/U(t)\to0$, and the auxiliary function is slowly
varying, so $a(pt)/a(t)\to1$. Writing $\phi(m)=U(m)+\gamma a(m)+o(a(m))$, the
$\gamma$-terms cancel to first order:
$\phi(\poolsize)-\phi(p\poolsize)
=U(\poolsize)-U(p\poolsize)+\gamma\bigl[a(\poolsize)-a(p\poolsize)\bigr]
+o(a(\poolsize))=a(\poolsize)\ln(1/p)\,(1+o(1))$, hence
\[
\loss(p)=\frac{\phi(\poolsize)-\phi(p\poolsize)}{\phi(\poolsize)}
=\frac{a(\poolsize)}{U(\poolsize)}\,\ln(1/p)\,(1+o(1)),
\]
so \emph{every} infinite-endpoint Gumbel-domain law is logarithmic in $1/p$
at fixed $p$, with platform-size relief governed by
$a(\poolsize)/U(\poolsize)$: $1/\ln\poolsize$ for the exponential,
$1/(2\ln\poolsize)$ for the Gaussian (there $U(t)\sim\sqrt{2\ln t}$, so
$a(t)=t\,U'(t)\sim1/\sqrt{2\ln t}$), and $\sigma/\sqrt{2\ln\poolsize}$ for
the lognormal, matching \Cref{eq:lognormal}. This is the general form of
\Cref{rem:gumbel}.

\subsection{Proof of \texorpdfstring{\Cref{thm:tail}}{Theorem 1}}
We compute $\phi(m)$ in each regularly varying domain and substitute into
$\loss(p)=1-\phi(p\poolsize)/\phi(\poolsize)$.

\paragraph{(1) Heavy tail, Fr\'echet, $\bar F(x)=x^{-\alpha}$, $x\ge1$, $\alpha>1$.}
With $a_m=m^{1/\alpha}$,
\[
\Prob\!\big(M_m\le a_m x\big)=\big(1-(a_m x)^{-\alpha}\big)^{m}
=\Big(1-\tfrac{x^{-\alpha}}{m}\Big)^{m}\to e^{-x^{-\alpha}},
\]
so $M_m/a_m\Rightarrow \Phi_\alpha$ (Fr\'echet), whose mean is
$\E[\Phi_\alpha]=\Gamma(1-1/\alpha)$ for $\alpha>1$. Choosing $r\in(1,\alpha)$,
$\sup_m\E[(M_m/a_m)^{r}]<\infty$, so $\{M_m/a_m\}$ is uniformly integrable and
$\phi(m)=\Gamma(1-1/\alpha)\,m^{1/\alpha}(1+o(1))$. Hence
\[
\loss(p)=1-\frac{(p\poolsize)^{1/\alpha}}{\poolsize^{1/\alpha}}(1+o(1))
=1-p^{1/\alpha}(1+o(1)),
\]
which is \Cref{eq:frechet}; note the independence from $\poolsize$. For a general
slowly varying $\ell$, the norming becomes $a_m=m^{1/\alpha}\ell^{\#}(m)$ with
$\ell^{\#}$ slowly varying, and
$\phi(p\poolsize)/\phi(\poolsize)
=p^{1/\alpha}\,\ell^{\#}(p\poolsize)/\ell^{\#}(\poolsize)\,(1+o(1))
\to p^{1/\alpha}$:
for \emph{fixed} $p\in(0,1)$, $\ell^{\#}(p\poolsize)/\ell^{\#}(\poolsize)\to1$ by
the uniform convergence theorem for slowly varying functions (uniformity on
compact subsets of $(0,\infty)$); the loss law is unchanged. Uniform
integrability for general $\ell$ follows from Potter bounds:
$m\bar F(a_m x)\le C x^{-\alpha+\delta}$ for large $m$ and any
$0<\delta<\alpha-1$, giving the integrable envelope
$\Prob(M_m>a_m x)\le\min\{1,Cx^{-\alpha+\delta}\}$ and hence moment
convergence~\cite{dehaan2006evt}. The fixed-$p$
restriction matters: if $p\to0$ jointly with $\poolsize\to\infty$, the ratio need
not converge.

\paragraph{(2) Bounded support, Weibull domain.}
Let $F$ have finite right endpoint $u^\ast$ with
$\bar F(u^\ast-t)=c\,t^{\alpha}(1+o(1))$ as $t\downarrow0$, $\alpha>0$. With
$a_m=(cm)^{-1/\alpha}$,
\[
\Prob\!\big(M_m\le u^\ast-a_m x\big)=\big(1-c(a_m x)^{\alpha}(1+o(1))\big)^{m}
\to e^{-x^{\alpha}},
\]
the reversed Weibull, so $(u^\ast-M_m)/a_m\Rightarrow$ a law with mean
$\Gamma(1+1/\alpha)$. The endpoint condition supplies an integrable envelope:
for $0\le x\le t_0/a_m$ (with $t_0$ small enough that
$\bar F(u^\ast-t)\ge c\,t^{\alpha}/2$ on $(0,t_0]$),
$\Prob\bigl((u^\ast-M_m)/a_m>x\bigr)=F(u^\ast-a_m x)^m
\le e^{-m\bar F(u^\ast-a_m x)}\le e^{-x^{\alpha}/2}$,
while the range $x>t_0/a_m$, of length $O(a_m^{-1})$, carries mass at most
$e^{-m\bar F(u^\ast-t_0)}$, whose geometric decay dominates that polynomial
growth; moment
convergence then gives
$\phi(m)=u^\ast-c'\,m^{-1/\alpha}(1+o(1))$ with $c'=\Gamma(1+1/\alpha)c^{-1/\alpha}$.
Thus
\[
\loss(p)=\frac{c'\big[(p\poolsize)^{-1/\alpha}-\poolsize^{-1/\alpha}\big]}
{u^\ast-c'\poolsize^{-1/\alpha}}
=\frac{c'}{u^\ast}\poolsize^{-1/\alpha}\big(p^{-1/\alpha}-1\big)(1+o(1)),
\]
which is \Cref{eq:weibull}. \qed

\subsection{Proof of \texorpdfstring{\Cref{prop:slate}}{Proposition 2}}
For fixed $j$, the $j$-th largest of $m$ draws satisfies
$(\util^{(j)}_{(m)}-b_m)/a_m\Rightarrow$ a limit law with the \emph{same} norming
constants $(a_m,b_m)$ as the maximum (only the limiting distribution changes, via
its finite mean $\mu_{\xi,j}$)~\cite{dehaan2006evt}. Summing term-wise, with
moment convergence for each of the top-$k$ order statistics~\cite{dehaan2006evt},
$\E\big[\sum_{j=1}^{k}\util^{(j)}_{(m)}\big]=k\,b_m+a_m\sum_{j=1}^{k}\mu_{\xi,j}+o(a_m)$.
Substituting into $\loss_k(p)=1-\tfrac{k b_{p\poolsize}+a_{p\poolsize}\bar\mu}{k b_{\poolsize}+a_{\poolsize}\bar\mu}$
($\bar\mu=\sum_j\mu_{\xi,j}$) and using the domain-specific $(a_m,b_m)$ reproduces
\Cref{eq:gumbel,eq:lognormal,eq:frechet,eq:weibull} up to constants: the $p$-
and $\poolsize$-dependence is unchanged. \qed

\subsection{Proof of \texorpdfstring{\Cref{prop:power}}{Proposition 3}}
Each arm's mean over $\approx p|\viewers|$ viewers with negligible
cross-viewer covariance has variance $\sigma^2(p)/(p|\viewers|)$; the
difference of two independent arms doubles it, and
$\mathrm{MDE}\propto\sqrt{\mathrm{Var}}$. With nonnegligible within-arm
covariance (e.g.\ coupling through shared in-arm supply), the variance is
inflated by the usual design-effect factor. \qed

\subsection{Reference-relative loss}
\label{app:refrel}
Experiments often measure loss against a partial reference rather than the
full catalog. For a reference retaining fraction $p_0$ with $0<p\le p_0\le1$ (both fixed as $\poolsize\to\infty$), define
\[
\loss(p;p_0)\eqdef1-\frac{\phi(p\poolsize)}{\phi(p_0\poolsize)}
=\frac{\loss(p)-\loss(p_0)}{1-\loss(p_0)}.
\]
Equivalently, every loss law above applies under the substitution
$\poolsize\mapsto p_0\poolsize$, $p\mapsto p/p_0$; for the heavy tail,
$\loss(p;p_0)=1-(p/p_0)^{1/\alpha}+o(1)$.

\section{Derivations of the traffic floors}
\label{app:floors}

Each row of \Cref{eq:tolfloor} inverts its loss law at tolerance
$\varepsilon$; we record the calculations, the constants they carry, and the
proof of \Cref{cor:scaling}.

\paragraph{Exponential.}
Setting the leading term of \Cref{eq:gumbel} equal to $\varepsilon$ gives
$\ln(1/p_\varepsilon)=\varepsilon(\ln\poolsize+\gamma)$, i.e.\
$\pmin_\varepsilon=e^{-\gamma\varepsilon}\poolsize^{-\varepsilon}$.

\paragraph{Lognormal.}
By the proof of \Cref{prop:exp}(ii),
$\loss(p)=1-e^{\sigma(z_{p\poolsize}-z_\poolsize)}(1+o(1))$, so the tolerance
binds when
$\sigma(z_{p_\varepsilon\poolsize}-z_\poolsize)=\ln(1-\varepsilon)$, i.e.\
$z_{p_\varepsilon\poolsize}=z_\poolsize-d_\varepsilon$ with
$d_\varepsilon=-\ln(1-\varepsilon)/\sigma>0$. Here
$p_\varepsilon=p_\varepsilon(\poolsize)\to0$, but
$z_{p_\varepsilon\poolsize}=z_\poolsize-O(1)$, a regime in which the
lognormal moment expansion above is uniform. It also implies
$\ln(p_\varepsilon\poolsize)/\ln\poolsize\to1$, so the $\ln\ln$ terms in
$z_m^2=2\ln m-\ln\ln m-\ln 4\pi+o(1)$ differ by $o(1)$. Thus
$2\ln p_\varepsilon=z_{p_\varepsilon\poolsize}^2-z_\poolsize^2+o(1)
=-2z_\poolsize d_\varepsilon+d_\varepsilon^2+o(1)$, hence
\[
\pmin_\varepsilon=(1-\varepsilon)^{\sqrt{2\ln\poolsize}/\sigma}\,
e^{d_\varepsilon^2/2}\,(1+o(1)):
\]
the row of \Cref{eq:tolfloor} keeps the leading factor, and
$e^{d_\varepsilon^2/2}$ is precisely the omitted constant noted in
\Cref{sec:scaling}.

\paragraph{Heavy tail.}
$1-p^{1/\alpha}=\varepsilon$ gives
$\pmin_\varepsilon=(1-\varepsilon)^{\alpha}$; $\poolsize$ does not enter.

\paragraph{Bounded.}
Setting $(c'/u^\ast)\poolsize^{-1/\alpha}(p^{-1/\alpha}-1)=\varepsilon$ and
solving gives
$\pmin_\varepsilon=\bigl(1+\tfrac{\varepsilon u^\ast}{c'}\poolsize^{1/\alpha}\bigr)^{-\alpha}$;
as $\poolsize\to\infty$,
$\pmin_\varepsilon\poolsize\to(c'/(\varepsilon u^\ast))^{\alpha}=O(1)$, which
is why this row is a small-$\varepsilon$ approximation and the exact floor
follows by inverting $\phi$ directly (\Cref{sec:scaling}).

\paragraph{Proof of \texorpdfstring{\Cref{cor:scaling}}{Corollary 1}.}
Write $\poolsize=\kappa\nactive$ with $\kappa$ fixed. The supply floor is
$\tilde m/\nactive\propto1/\nactive$. Bounded:
$\pmin_\varepsilon\sim(c'/(\varepsilon u^\ast))^{\alpha}\poolsize^{-1}
\propto1/\nactive$, the same rate, so
$\pmin\propto1/\nactive$. Exponential-type:
$\pmin_\varepsilon\propto\poolsize^{-\varepsilon}$ decays slower than
$1/\nactive$ for $\varepsilon<1$, so it binds at scale and
$\pmin=\nactive^{-\varepsilon+o(1)}$ (slowly varying factors do not alter the
exponent). Lognormal:
$\pmin_\varepsilon=e^{-d_\varepsilon\sqrt{2\ln\poolsize}}(1+o(1))$ up to the
constant above, slower than any power of $\nactive$. Heavy:
$\pmin_\varepsilon=(1-\varepsilon)^{\alpha}$ is constant in $\nactive$, so
only the supply floor improves with scale. Taking the maximum in each case
gives the corollary. \qed

\paragraph{Anchor robustness.}
Replacing $\util$ by $\util+\kappa_0$ for a constant $\kappa_0>0$ (a location
shift of the engagement anchor of \Cref{sec:model-setup}) preserves each
max-domain and the orders in $p$ and $\poolsize$ of every law above. In the
Gumbel family, $b_m\mapsto b_m+\kappa_0$, so
$\loss(p)=\tfrac{b_\poolsize-b_{p\poolsize}}{b_\poolsize+\kappa_0}(1+o(1))$
and, since $b_\poolsize\to\infty$, the relative change in fixed-$p$ loss is
only $O(\kappa_0/b_\poolsize)$. For a regularly varying tail,
$\Prob(\util+\kappa_0>x)=\bar F(x-\kappa_0)=x^{-\alpha}\tilde\ell(x)$ with
$\tilde\ell$ slowly varying, so \Cref{eq:frechet} is unchanged (it depends on
$F$ only through $\alpha$). In the bounded case the endpoint moves to
$u^\ast+\kappa_0$, and only the constant $c'/u^\ast$ in \Cref{eq:weibull}
(hence in the bounded row above) changes. The floors' $\poolsize$-scaling is
therefore anchor-robust, although finite-sample and multiplicative constants
can depend on the anchored scale.

\section{Toward a creator-side dual}
\label{app:dual}

\Cref{rem:dual} notes that per-creator statements need a model of how the
recommender allocates viewers across competing items. The mean does not.

\paragraph{The mean dual is allocation-free.}
Consider a symmetric cell at fraction $p$ ($pV$ viewers with
$V=|\viewers|$, $p\nactive$ creators, and $p\poolsize$ items by
\Cref{lem:thin}), each viewer issuing
requests at a common rate $R$, each request served the in-arm argmax as in
\Cref{def:model}. Total expected engagement in the cell is
$pV\!R\,\phi(p\poolsize)$; dividing by the $p\nactive$ creators, the
\emph{mean} received engagement per creator is
$\bar E(p)=(V\!R/\nactive)\,\phi(p\poolsize)$, and its ratio to the full-platform mean
is $\phi(p\poolsize)/\phi(\poolsize)=1-\loss(p)$: the average creator-side
loss obeys the same law as the viewer-side loss, for any tail class; the
statement is exact for a deterministic pool and asymptotic under the
creator-block randomization and vanishing-largest-share condition of
\Cref{lem:thin}. It uses only conservation of total engagement; how the loss
distributes \emph{across} creators is what requires an allocation model.

\paragraph{Matched cells and the conservation check.}
For two matched-fraction cells $p_1<p_2$ the identity gives
\[
\bar E(p_1)/\bar E(p_2)=\phi(p_1\poolsize)/\phi(p_2\poolsize)
=1-\loss(p_1;p_2),
\]
the reference-relative loss of Appendix~\ref{app:refrel}: with means taken
over all assigned creators, conservation makes the creator-side and
viewer-side relative losses coincide. The production creator metric
conditions on creators with at least one submission, so the comparison is a
coherence check rather than an exact identity. The data exhibit this
approximate accounting relationship: between
the $2\%\times2\%$ and $10\%\times10\%$ cells, mean received exploration
view time per creator is $20.9\%$ lower while per-viewer exploration view
time is $19.9\%$ lower (\Cref{tab:sweep}); this agreement is implied by
conservation rather than independent validation.

\paragraph{Upper-extreme reach under an illustrative allocation.}
As one explicit model (beyond the paper's assumptions), give item $i$ a
quality weight $w_i$ and give viewers Gumbel taste shocks, so a request lands on
item $i$ with the Luce probability $w_i/\sum_j w_j$ over the in-arm items.
A retained item's expected \emph{reach} (requests won) is then
$pV\!R\,w_i/(p\poolsize\,\E[w])=(V\!R/\poolsize)\,(w_i/\E[w])(1+o(1))$
for weights with $\E[w]<\infty$: the per-item reach distribution is
asymptotically invariant under thinning, while mean received
\emph{engagement} still falls by $\loss(p)$ through the quality of the
matches won. Only the cell's extremes move: for regularly varying weights
with index $\alpha_w>1$, the largest reach in the cell is
$p^{1/\alpha_w}(1+o(1))$ times its full-pool counterpart, a relative
reduction of $1-p^{1/\alpha_w}+o(1)$, the same functional form as
\Cref{eq:frechet}. This is an extreme-order statement, not a prediction
about fixed reach tiers.

\paragraph{Scope.}
The raw creator-side estimates in \Cref{sec:empirical} are reported against
the $10\%$-creator, $70\%$-audience reference, so they also reflect audience
share and production impression pacing and are not themselves estimates of
$\loss(p)$; only their matched-cell ratio enters the conservation check
above.

\section{Additional simulation panels}
\label{app:steepening}

\begin{figure}[t]
\centering
\includegraphics[width=0.8\linewidth]{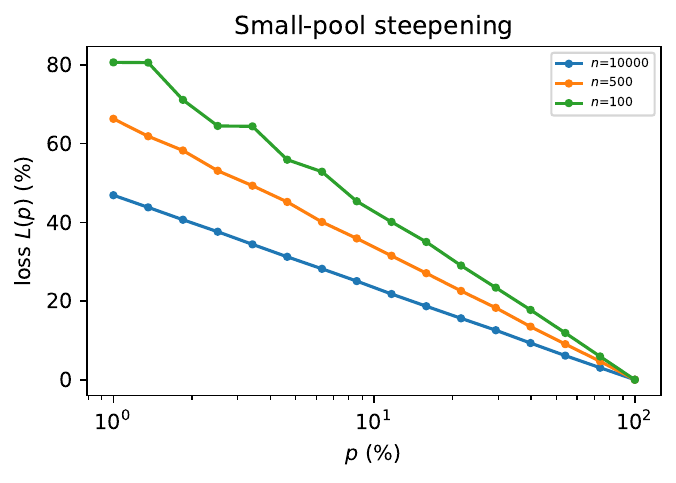}
\caption{Shrinking the pool from $n=10^4$ to $n=10^2$ steepens the
exponential loss curve: the model's own small-pool curvature, illustrating
(not validating) behavior near the supply floor of \Cref{sec:scaling}; the
simulation contains no scarcity or backfill mechanics.}
\Description{Loss versus isolation fraction for three pool sizes; smaller
pools show steeper loss curves.}
\label{fig:steepening}
\end{figure}

\begin{figure}[t]
\centering
\includegraphics[width=0.8\linewidth]{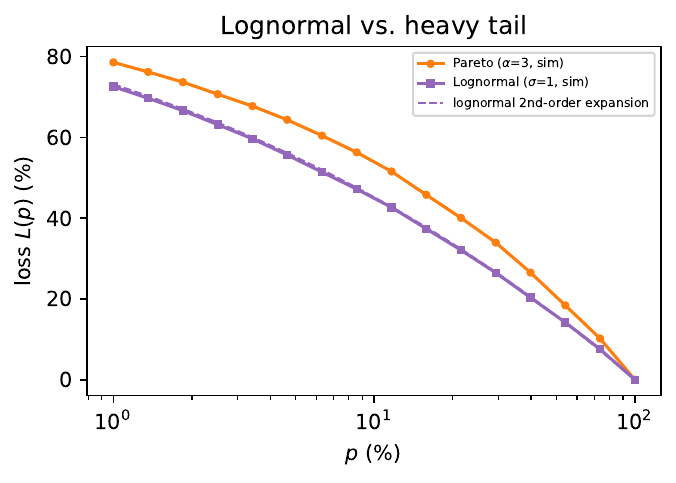}
\caption{The simulated lognormal ($\sigma=1$) parallels the heavy-tail curve
(Pareto, $\alpha=3$) in shape across the simulable range: the power-law versus
lognormal ambiguity of \Cref{rem:gumbel}, made visible. The overlay is the
second-order expansion from the proof of \Cref{prop:exp}(ii); the first-order
law \Cref{eq:lognormal} converges only at $1/\sqrt{\ln n}$ rates and is not a
good description at simulable $n$.}
\Description{Loss versus isolation fraction: the lognormal curve lies close
to the Pareto curve, with a dashed second-order expansion matching the
lognormal points.}
\label{fig:lognormal-sim}
\end{figure}

\Cref{fig:steepening} illustrates the model's behavior as the pool approaches
the supply-limited boundary of \Cref{sec:regimes}; \Cref{fig:lognormal-sim}
shows that the lognormal and heavy-tail losses are qualitatively similar over
the simulated range.

\section{Tail diagnostics on production marginals}
  \label{app:tails}

\begin{figure}[t]
\centering
\includegraphics[width=0.92\columnwidth]{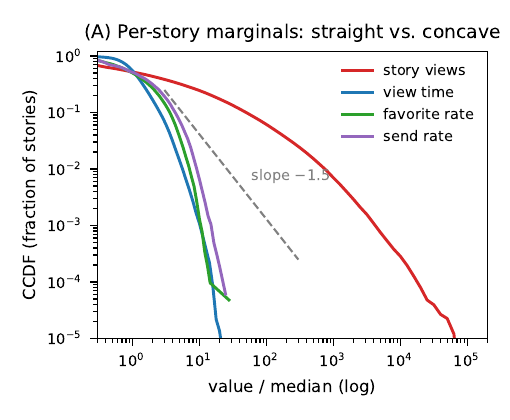}
\caption{Tail diagnostics on production per-story marginals (axes normalized
by per-metric medians; absolute scales withheld): popularity is scale-free
(straight, slope $\approx\!-1.5$), while deep engagement metrics are concave
with steepening local slope, a lognormal-like signature.}
\Description{Complementary cumulative distributions on log-log axes: story
views form a straight line while view time, favorite, and send curves bend
downward.}
\label{fig:tails}
\end{figure}

\begin{figure}[t]
\centering
\includegraphics[width=0.92\columnwidth]{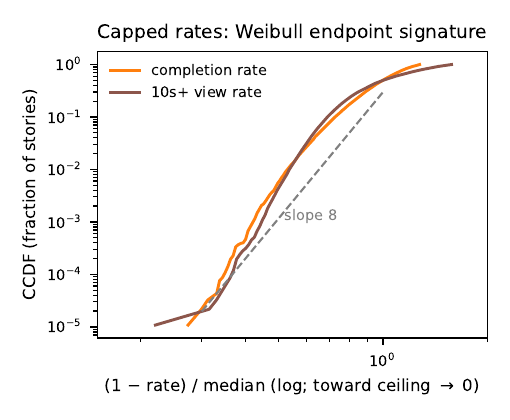}
\caption{Capped rates against distance to their ceiling (axis normalized by
the median of $1-\text{rate}$, which leaves the slope unchanged): straight
decay of slope $\approx\!8$--$10$, the Weibull endpoint signature.}
\Description{Completion and ten-second view rates fall steeply and linearly
against one minus rate on log-log axes, indicating a bounded endpoint.}
\label{fig:tails-b}
\end{figure}

  \Cref{fig:tails,fig:tails-b} plot per-story marginals of the \Cref{tab:sweep} metrics
  over a recent snapshot of the exploration corpus as log--log complementary
  cumulative distribution functions (CCDFs). For platform confidentiality,
  axes are normalized by per-metric medians (in \Cref{fig:tails-b}, by the
  median of $1-\text{rate}$, which leaves the endpoint slope unchanged); only
  distributional shape is disclosed. These marginals average each story over
  its viewers and condition on the production ranker having served it, so they
  do not estimate the per-viewer match quality $F$ and cannot identify its
  max-domain. Averaging generally compresses dispersion, while serving
  selection truncates or reweights the observed range; we therefore read these
  marginals only as consistency checks. Three shapes emerge, by
  metric type. Story views (popularity) are scale-free over the observable
  range (log--log slope $\approx\!-1.5$), echoing the popularity-skew driver of
  \Cref{sec:tail-law}; as an exposure count rather than a match-quality
  proxy, they are not themselves evidence on $F$. The engagement metrics far
  from any cap (view time, favorite, send) are concave, with local slope steepening from
  $\approx\!3$--$4$ to $\approx\!6$--$7$ at the deepest measurable quantiles: a
  lognormal-like signature whose effective index trends toward the calibrated
  $\alpha\approx9$--$15$, probed at the still-deeper quantiles sampled by
  in-arm maxima. The capped
  rates (completion, $10^{+}$s view rate), measured near their soft ceilings, decay
  with the Weibull endpoint signature at index $\approx\!8$--$10$ (\Cref{fig:tails-b}).
  None of the engagement marginals is decisively Fr\'echet and none
  contradicts the effective-heavy calibration; over the operating range the classes are indistinguishable
  (\Cref{rem:gumbel}), which is why \Cref{sec:empirical} calibrates rather
  than identifies.

\end{document}